\documentclass[preprint,12pt]{elsarticle}

\usepackage{color}
\usepackage{amssymb}
\usepackage{booktabs} 
\usepackage[utf8]{inputenc}
\usepackage[english]{babel}
\usepackage{amsfonts}
\usepackage{amsmath}
\usepackage{array}
\usepackage{epsf}
\usepackage{textgreek}
\usepackage{epsfig}
\usepackage[version=3]{mhchem}
\usepackage{textcomp}
\usepackage{graphics}
\usepackage{graphicx}
\usepackage{epstopdf}
\newcommand{\Fig}[1]{Fig.~\ref{#1}}

\journal{Chemical Physics}

\begin{document}

\begin{frontmatter}

\title{DFT-D Investigation of the Interaction Between	
Ir(III) Based Photosensitizers and Small Silver Clusters Ag$_n$ ($n$=2--20, 92)}

\author[ur]{Olga S. Bokareva \corref{cor1}}
\author[ur]{Oliver K\"uhn}
\address[ur]{Institut f\"{u}r Physik, Universit\"{a}t Rostock, D-18051 Rostock, Germany}

\cortext[cor1]{Email: obokareva@gmail.com }		

\begin{abstract}
A dispersion-corrected density functional theory 
study of the photosensitizer [Ir(ppy)$_2$(bpy)]$^+$
and its derivatives bound to silver clusters Ag$_n$ ($n$=2--20, 92)
is performed. The goal is to provide a new system-specific
 set of $C_{\rm 6}^{}$ interaction  parameters for 
 Ag and Ir atoms. To this end a  QM:QM scheme is 
 employed using the PBE functional and RPA as well as MP2 calculations as 
 reference.
The obtained $C_{\rm 6}^{}$ coefficients were applied to determine
dissociation curves of selected \ce{IrPS-Ag_{$n$}} 
complexes and binding energies of
derivatives containing oxygen and
sulphur as heteroatoms in the ligands. 
Comparing different $C_{\rm 6}^{}$ parameters it is
concluded that RPA-based dispersion correction produces binding energies close to standard D2 
and D3 models, whereas MP2-derived parameters overestimate these energies.

\end{abstract}
\begin{keyword}
	dispersion interaction \sep density functional theory \sep
	  organic/inorganic hybrid systems \sep binding energies
\end{keyword}
\end{frontmatter}

%
\section{Introduction}
\label{sec:intro}
%
Inorganic/organic hybrid systems comprised of small metal 
nanoparticles and different organic adsorbates 
like peptides and dyes represent a fascinating topic
with prospective applications in, e.g.,  catalysis and bioelectronics
(for reviews, see Refs.
\cite{Dulkeith-prl-2002,Anger-prl-2006,Pustovit-prb-2011,Zweigle-jpcc-2011}).

Despite considerable progress, electronic structure calculations
of such hybrid systems with non-covalent interactions
still pose a challenge for quantum chemical 
methods that are known for  their moderate computational costs. 
In particular, standard density functional theory
(DFT) techniques fail to predict 
the adsorption energies for aromatic molecules 
on noble metal surfaces in agreement with 
experiments \cite{Bilic-jpcb-2002,Bilic-jctc-2006,Higai-ss-2006}. 
Often this discrepancy  can be attributed
to the inadequate treatment of the 
dispersion interaction within most currently available 
functionals \cite{Kristyan-cpl-1994,Hobza-jcc-1995,
Perez-cpl-1995,Perez-jcp-1999,Grimme-cms-2011}.
This holds true in particular for  weakly physisorbed 
adsorbates, where the incorrect estimation of dispersion 
interaction might have a much more pronounced impact
on the quality of results than in case of covalently 
bound complexes \cite{Bilic-jpcb-2002,Bilic-jctc-2006,Higai-ss-2006}. 

\textcolor{black}{Since the dispersion interaction is the main attractive force 
holding weakly bound molecules together, it stands in modern literature
for the more general van der Waals
(vdW) interaction, which includes different types of intermolecular interactions such as 
electrostatic (permanent multipole-permanent multipole), 
induction (permanent multipole-induced multipole), 
and dispersion (induced multipole-induced multipole).
Strictly speaking, only many-electron quantum chemistry methods can reliably
describe the dispersion term, as it arises from the correlated motion of electrons.
Hence, for computational reasons, the direct calculation of vdW interactions is limited
to small systems. That is why alternative approximate approaches 
to the  investigation of weakly-bounded systems at reasonable computational
costs are warrant and }
the extension of standard DFT to account for dispersion 
effects is an active area of research (for a review, see 
Refs. \cite{Tkachenko-mrsb-2010, Grimme-cms-2011, Klimevs-jcp-2012}). 
\textcolor{black}{Here, the strategies can be classified according to the 
basis on which the dispersion term is founded, i.e. the effective one electron potential,
the interaction between pairs of atoms with $C_{6}^{}$ 
atomic coefficients or the 
inclusion of non-local terms in the exchange-correlation (XC) kernel.}

\textcolor{black}{The first approach includes dispersion-corrected 
atom-centered potentials (DCACP) 
\cite{Lilienfeld-prl-2004, Lilienfeld-prb-2005} as well as its local variants LAP
\cite{Sun-jcp-2008} and CAP \cite{Johnson-jpoc-2009}
and the recently proposed empirical force correcting atom-centred potentials
(FCACP's) \cite{Lilienfeld-mp-2013}. It makes use of tuned 
effective core potentials placed at each atom in the way that the dispersion
energy is represented by a sum of one-electron terms.}

\textcolor{black}{The idea of representing the dispersion interaction by pairwise atomic
$-C_{n}^{}R^{-n}_{}$ potentials with proper 
damping functions was explored by many authors.
In the exchange-hole dipole moment (XDM) approach 
\cite{Becke-jcp-2005,Becke-jcp-2006,
Angyan-jpc-2007,Becke-jcp-2007,Hesselmann-jcp-2009,
Steinmann-jctc-2010,Hesselmann-jcp-2012}, 
the instantaneous dipole formed between an electron 
and its exchange hole is used to express the dispersion
interaction between non-overlapping charge densities using
the Casimir-Polder relation \cite{Casimir-pr-1948}. In most 
of the recent implementations, the pairwise potentials includes
terms up to tenth order ($C_{6}^{}$, $C_{8}^{}$ 
and $C_{10}^{}$ coefficients). The impact of the
chemical environment on the dispersion
coefficients is taken into account through Hirshfeld partitioning.}

In the DFT-D2 approach of Grimme, $C_{\rm 6}^{} $ parameters 
for the first part of the periodic table are 
\textcolor{black}{derived from calculations
of ionization potentials and static dipole polarizabilities 
for single atoms } and 
proved to give an adequate description of non-covalent 
interactions. 
For heavy atoms, however, no reliable
parameters within the D2 model are 
available \cite{Grimme-jcc-2006}. As 
proposed by Grimme, unknown coefficients 
could be deduced from atomic properties 
by the London formula \cite{Eisenschitz-zp-1930}.
An alternative way to obtain 
dispersion corrections is the so-called hybrid QM:QM 
approach \cite{Tuma-cpl-2004,Tuma-pccp-2006,Hu-prl-2007}. 
In this method, the dispersion energy 
is considered as the difference between the adsorption 
energies for adsorbate-substrate complex obtained 
with ab initio theory and DFT. For 
absorption of pyridine on gold clusters, it has been 
shown by Tonigold and Gro\ss{} that QM:QM 
employing a MP2 (second order M\o ller-Plesset 
perturbation theory) reference gave 
substantially better agreement with experiment
compared to standard D2 \cite{Tonigold-jcp-2010}.

Later, Grimme has introduced the much 
advanced DFT-D3 method, taking into 
account the surrounding of the atoms by 
means of coordination numbers in contrast to DFT-D2 where 
by construction the dispersion coefficients are
not system-dependent. In this approach,
$C_{\rm 6}^{}$ coefficients were derived from averaged dipole
polarizabilities at imaginary frequencies calculated 
with time-dependent DFT
(TD-DFT), while eighth order coefficients  $C_{\rm 8}^{} $ 
follow from a simple recursion rules for the
higher-multipole terms \cite{Grimme-jcp-2010}.
DFT-D3  has been shown
to provide higher accuracy and broader 
applicability for 94 elements of the periodic 
table than the earlier versions.

\textcolor{black}{Tkatchenko and Scheffler developed a method (TS or DFT-vdW) 
\cite{Tkachenko-prl-2009} of calculating dispersion
coefficients and vdW radii from the ground-state molecular
or condensed matter electron density.  Here, the starting point 
is a high-level ground-state calculation of free-atom properties.
In a second step the
effect of the surroundings 
 is taken into account by considering Hirshfeld
volumes and the electron density for the whole system. Further,
the self-consistent screening is accounted for to
reproduce the anisotropy of the molecular
static polarizability \cite{Tkatchenko-prl-2012}. In the DFT-vdWsurf variant
the many-body collective response is analyzed for 
cases of adsorbates on a surface \cite{Ruiz-prl-2012}. In that respect, the DFT+vdW
group is more akin to XDM, because it gives $C_{6}^{}$ coefficients
that are dependent on the local environment, but not only on coordination
number of each atom like in DFT-D3.}

\textcolor{black}{In passing we note that in order to avoid double counting effects at medium and short distances
a linkage between the long-range dispersion term 
and the short-range DFT energy must be introduced. In
all DFT+dispersion variants this is achieved by multiplying the
dispersion energy by a damping function that provides a smooth cut-off at
short distances. In general, this type of dispersion correction can by applied to different DFT variants, but
corresponding empirical coefficients should be properly calibrated.}

Finally, a more rigorous development 
\textcolor{black}{(noted as vdW-DFs)} sets the focus on the 
determination of an explicit non-local
correlation functional from first principles. 
\textcolor{black}{Currently, the following versions are used
vdW-DF \cite{Dion-prl-2004, Langreth-ijqc-2005, Lee-prb-2010}, VV09, and VV10
\cite{Vydrov-prl-2009, Vydrov-jcp-2010}.}
\textcolor{black}{The exchange-correlation energy includes a non-local term,
which is undamped (in modern versions) and contributes at short distances as well.
The kernel used to compute the non-local term is typically based on a local
approximation to the dipole polarizability at imaginary frequencies. This results in
a long-range part of the dispersion energy through the Casimir-Polder relation \cite{Casimir-pr-1948}, 
similar to modern DFT-D approaches to} $C_{6}^{}$ \textcolor{black}{coefficients. The similarity between
vdW-DF and DFT-D can also be seen from the fact, that the non-local term of the former approach is
typically computed non-self-consistently, resulting in some additional contribution to the DFT energy.
Both routes describe the long-range vdW interaction and, 
at atomic overlap regions, link it to a standard exchange-correlation functional.
The advantage of vdW-DF  is that dispersion effects are calculated based  on the charge
density, that is, in cases of  charge transfer the effect of  dispersion is naturally included.}
Recently, some progress in increasing 
their efficiency has been achieved
\textcolor{black}{making the computational costs 
comparable to those of standard DFT-GGA calculations}
what would enable their future application to large systems
\cite{Sato-jcp-2007,Roman-prl-2009, Gulans-prb-2009,Vydrov-jcp-2010}.
For example, vdW-inclusive DFT methods allow to reliably
model adsorption of molecules on surfaces
 \cite{Liu-prb-2012}, for review see \cite{Tkachenko-mrsb-2010}.

Although benchmarks for validation of new 
dis\-per\-sion-corrected DFT approaches 
have been reported (see, e.g., Refs. \cite{Steinmann-jctc-2010, Vydrov-pra-2010, 
Burns-jcp-2011,  Marom-jctc-2011, Steinmetz-co-2013,
Goerigk-jctc-2014}
and references therein), mainly non-metal
containing systems have been considered such as the
S22 standard set and nucleobase pairs. Exceptions 
include the application of DFT-D 
to the adsorption of aliphatic and  aromatic molecules on 
metal surfaces like gold, silver, palladium, and copper 
\cite{Tonigold-jcp-2010,Grimme-jcp-2010,Atodiresei-prl-2009,Atodiresei-prb-2008,
McNellis-prb-2009,Nguyen-pccp-2010,
Gaspari-prb-2010,McNellis-pccp-2010,
Antony-jcp-2012}, which 
gave a good agreement with  experimental 
adsorption energies,
in contrast to conventional DFT.
However, note that for AuL$_{x}^{}$-C$_{n}^{}$H$_{m}^{}$ 
model systems,  for instance, the DFT-D3 performance 
is comparable with that of conventional DFT  \cite{Kang-jctc-2011}.
An alternative for improving  the DFT-D performance 
is to use system-de\-pen\-dent
$C_{\rm 6}^{}$ coefficients, rescaled on the basis
of an embedding model; examples include 
the adsorption of small organic 
molecules on MgO and NaCl surfaces \cite{Ehrlich-cpc-2011}.
In conclusion, the usage of dispersion-corrected DFT 
approaches for systems including metal-containing
surfaces or \textcolor{black}{especially} clusters clearly needs further 
investigation and comprehensive benchmarking
in order to become a standard 
method comprising the modest computational 
costs of DFT with accurate predictions. 

\begin{figure}[t]
\begin{center}
\includegraphics[width=0.5\textwidth]{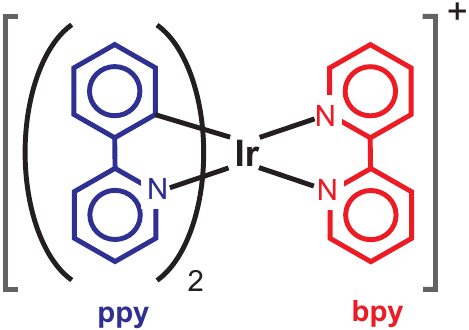}
\caption{Chemical formula of \ce{[Ir(ppy)2(bpy)]+}.}
\label{IrPSscheme}       
\end{center}
\end{figure}

The present study aims at establishing an 
empirical dispersion correction for the system 
\ce{[Ir(ppy)2(bpy)]+} (IrPS) 
shown in \Fig{IrPSscheme} and its derivatives, bound to 
small silver clusters Ag$_n$.
In doing so we will contrast the  conventional DFT-D technique, including 
D2, D3 corrections, with the hybrid  QM:QM approach for obtaining
\emph{problem-specific}  
$C_{\rm 6}^{}$ coefficients for the heavy atoms. 
Our choice of the system is motivated by
the use of Ir(III) complexes as photosensitizers in 
photocatalytic water splitting. 
In Ref. \cite{Gaertner-acie-2009}, for instance, 
the homogeneous catalytic system 
consisting of  IrPS combined with the sacrificial 
reductant  triethylamine and a water reduction
iron catalyst has been demonstrated. 
Hybrid systems consisting of IrPS and small metal clusters 
hold the promise to obtain a heterogeneous catalytic 
system with improved performance.

The interaction of IrPS with small silver clusters 
(1--6 silver atoms) and in particular
changes in absorption spectra upon binding
have been studied 
in Ref. \cite{Bokareva-pccp-2012} employing 
the long-range corrected DFT (LC-BLYP) 
approach. The obtained results demonstrated
strong changes in the absorption spectra 
of the combined systems as compared with the
pure constituents.  To proceed with larger
metal clusters  it is desirable to have a 
reliable method which properly describes binding 
interactions at low computational cost such as DFT-D.

The paper is organized as follows: First, we 
briefly recall the main features
of the DFT-D approach and outline
the computational details in Section 
\ref{sec:Comp}. Second, in Section \ref{sec:Res}
we present the results of fitting $C_{\rm 6}^{}$ 
coefficients for Ir and Ag atoms employing the
QM:QM procedure. We proceed with 
the applications of the new coefficient set to IrPS 
derivatives. Final conclusions are given in Section \ref{sec:Conc}.
%
\section{Computational Details}
\label{sec:Comp}
%
Initial geometry optimizations of \ce{IrPS-Ag_{$n$}}, 
\ce{bpy-Ag_{$n$}}, and \ce{ppyH-Ag_{$n$}}
complexes as well as of pure organic and metal parts 
were carried out using the generalized 
gradient approximation (GGA) functional of Perdew, 
Burke, and Ernzerhof (PBE) 
\cite{Perdew-prl-1996} along with the
def2-SV(P) basis set
\cite{Schaefer-jcp-1992,Eichkorn-tca-1997}.
Optimizations were carried out without 
any symmetry constraints. 
Starting geometries of small silver 
clusters Ag$_n$, $n$=2--20 were 
taken from previous studies \cite{Bonacic-jcp-1993,
Bonacic-jcp-2001,Fournier-jcp-2001,Harb-jcp-2008,Baishya-prb-2008}. 
On these optimized geometries, single 
point calculations with PBE, Random Phase Approximation (RPA), and MP2
were done employing the def2-TZV(P) basis set 
\cite{Weigend-pccp-2005,Weigend-cpl-1998}.

The binding energy, $E_{\rm b}^{} $, has been defined as

\begin{equation} \label{Ebind} 
E_{\rm b}^{}=E_{\rm tot}^{}-E_{{\rm Ag}_{n}}^{} - E_{\rm IrPS}^{}
\end{equation}

where  $E_{\rm tot}^{} $, $E_{\rm IrPS}^{} $, 
and $E_{{\rm Ag}_{n}}^{} $ 
are the total energies of the relaxed complex, dye molecule,  
and silver cluster, respectively.

It goes without saying that at the moment for
the present system, for example, CCSD(T) reference calculations are out of
reach. Higher order perturbation theory (MP3, MP4) doesn't improve 
the situation as shown in Ref. \cite{Tonigold-jcp-2010}. 
An alternative is the RPA method, 
which is of slightly higher computational cost as MP2, see \cite{Eshuis-tca-2012}. 
Unlike MP2, the RPA method does not suffer from problems like infinite 
energies for small bandgap systems and it  was shown to provide an adequate description for
non-covalent interactions \cite{Eshuis-jcp-2012}. These authors also  pointed out 
that sufficiently accurate binding energies 
of weakly bound systems can be obtained only by
using complete basis set extrapolation or basis sets larger than 
quadruple-\textzeta. But, this level is hardly affordable 
for heavy elements like Ag or Ir, not to mention 
that besides regular basis sets one would  
need auxiliary ones to perform RI-calculations 
\cite{Eshuis-jcp-2010,Eshuis-tca-2012}. Therefore, 
we are forced to restrict our considerations to basis sets of triple-\textzeta\ quality.
At least for the case of MP2 we have performed a convergence study and concluded that def2-TZVP binding energies are almost
saturated with respect to basis set; e.g. the corresponding
binding energy in MP2 of \ce{IrPS-Ag2} were -0.887, -1.000, and -0.961 eV
for def2-SV(P), def2-TZVP, and def2-QZVP, respectively.

For the RPA calculations we have employed the resolution-of-the-identity 
approximation (RI-RPA) for the two-electron  integrals \cite{Weigend-pccp-2002,Sierka-jcp-2003,Weigend-tca-1997,
Haettig-jcp-2000,Haettig-pccp-2006},
the corresponding auxiliary basis sets were taken 
from Refs. \cite{Eichkorn-tca-1997,Eichkorn-cpl-1995,
Weigend-cpl-1998,Hellweg-tca-2007}. 
The RI-RPA calculations \textcolor{black}{of the correlation energy} were done 
\textcolor{black}{non-self-consistently} on top of the PBE 
converged set of Kohn-Sham molecular orbitals with 
the number of grid points set equal to 30.
\textcolor{black}{In terms of the notation introduced in \cite{Ren-prl-2011},
the cRPA@PBE approach is applied in this work. For
simplicity, we will further use just the abbreviation RPA. }
In the RPA and MP2 calculations, 
the frozen-core approximation
was used to speed up the calculations and to avoid the using basis sets
with additional tight core correlation functions.
According to default selection, 
orbitals below -3 $E_{\rm Hartree}$ were frozen, i.e. 
1s of nitrogen and carbon atoms,
4s of silver and 5s of iridium. All calculations were  done 
with the TURBOMOLE 6.3 program package \cite{TURBOMOLE}.
 
We introduced the dispersion term correcting 
DFT results according to the Grimme DFT-D2 
model \cite{Grimme-jcc-2004,Grimme-jcc-2006}. Here
the total energy of system is given by expression

\begin{equation} \label{EDFTD}
E_{\rm DFT-D}^{} = E_{\rm DFT}^{} + E_{\rm disp}^{} \, ,
\end{equation}

where $E_{\rm DFT}^{}$ is the energy obtained from 
the DFT calculation and $E_{\rm disp}^{}$ is a dispersion term
including the $C_{\rm 6}^{} R^{\rm -6}_{}$ dependence. 
$E_{\rm disp}^{}$ has been determined as introduced by 
Grimme 
\textcolor{black}{for DFT-D2} \cite{Grimme-jcc-2004,Grimme-jcc-2006}

\begin{equation} \label{Edisp}
E_{\rm disp}^{}=-s_{\rm 6}^{}\sum_{i}\sum_{j}\frac{C_{\rm 6}^{ij}}
{R_{ij}^{6}}f_{\rm damp}^{}(R_{ij}^{})
\end{equation}

\begin{equation} \label{fdamp}
f_{\rm damp}(R_{ij})=\frac{1}{1+\exp[-d(R_{ij}/R_{r}-1)]}
 \end{equation}

where $R_{\rm r}^{}$ is the sum of van der Waals radii of the interacting
atoms, $d$ determines the steepness
of the damping function, and $C_{\rm 6}^{}$ is obtained as 

\begin{equation} \label{C6ij}
C_{\rm 6}^{ij}=\sqrt{C_{\rm 6}^{i}C_{\rm 6}^{j}} \, .
\end{equation}

The damping function $f_{\rm damp}^{}$, the scaling 
factor $s_{\rm 6}^{}$, and the atomic $C_{\rm 6}^{i}$ 
coefficients for non-metal atoms have been taken without changes 
from Ref. \cite{Grimme-jcc-2006}. 

Atomic coefficients  for metal atoms (silver and iridium) 
have been obtained using the hybrid QM:QM me\-thod
proposed in Ref. \cite{Tuma-cpl-2004} and
applied to metal surfaces 
in Ref. \cite{Hu-prl-2007}. In this approach, 
we assume the dispersion energy 
to be the energy difference between
binding energies calculated with 
the reference (RPA or MP2) and DFT 
(for discussion see Section \ref{sec:Res}C). 
The application of quite large basis 
sets allows one to neglect
the  BSSE correction as it was pointed out
by Grimme \cite{Grimme-jcc-2006}.
Differences in the binding energies between MP2/RPA and  PBE for
different numbers of silver atoms have been least-square fitted 
using Eq. (\ref{Edisp}) in order to obtain atomic 
coefficients. 

Because of the fact that the target systems 
(IrPS-Ag$_n$) include two metal elements the task was divided 
into two  steps. First, we considered structures 
consisting of the same small silver clusters 
and phenylpyridin (ppyH) or bi\-py\-ri\-din 
(bpy) molecules. For these cases, we 
only needed to 
approximate the dispersion coefficient for 
silver. This has been done using MP2 and RPA references. Second, we regarded the 
set of \ce{IrPS-Ag_{$n$}} structures
 and 
did the same fitting procedure 
applying the coefficient for silver 
calculated at the first stage. However, for computational reasons this was possible for the MP2 reference only.
Alternative to this two step procedure we approximated the 
two coefficients simultaneously  using all dependencies. In this respect, the 
addition of sets of model structures,
including separate ligands as organic 
part, was reasonable
because the interaction between 
IrPS and silver clusters
is mainly due to the dispersion 
interaction with ligands,
with the central Ir atom being shielded.

%
\section{Results and discussion}
\label{sec:Res}
\subsection{Geometries of Weakly Bound Complexes}
\begin{figure}[t]
\begin{center}
\includegraphics[width=0.7\textwidth]{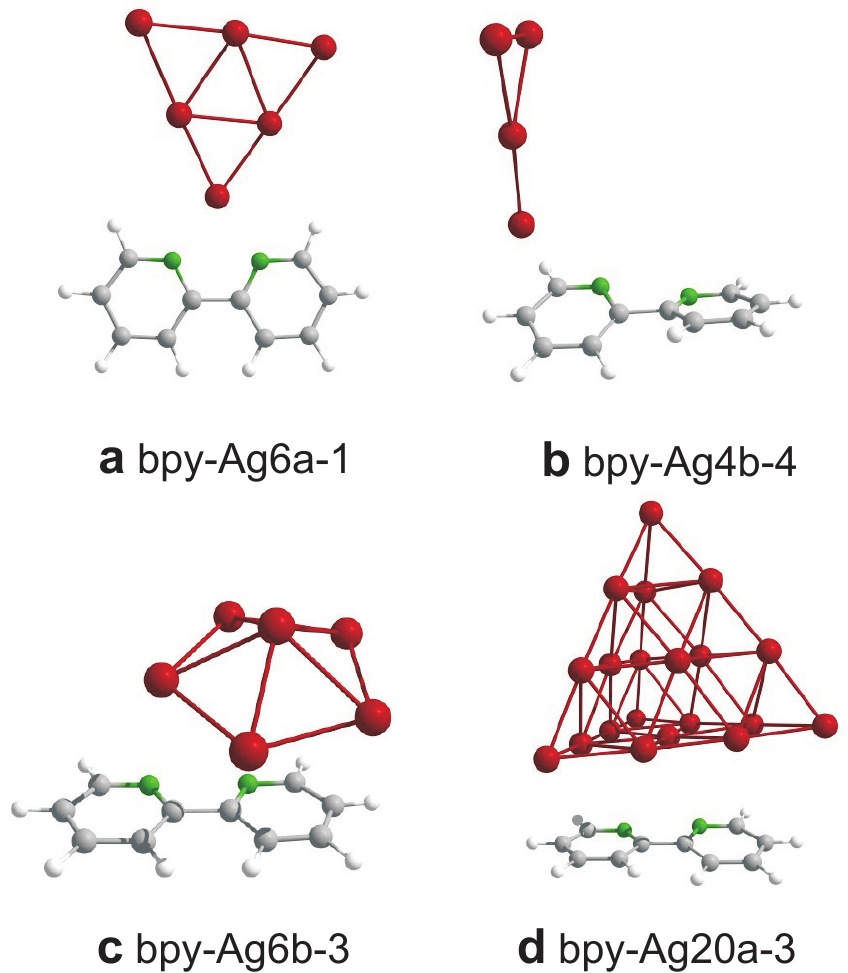}
\end{center}
\caption{\label{bpy-Agn-examples}
Some representative examples of \ce{bpy-Ag_{$n$}}
optimized geometries (for a full list 
including the nomenclature see Supplementary Material).}
\end{figure}

%
For each given combination
of silver cluster and organic molecule,
we have first optimized from 2 to 5 different
structures starting from various
initial relative orientations. 
For brevity, in  \Fig{bpy-Agn-examples} only
some examples  of optimized complexes are plotted, 
for the full list of structures and their notation see 
the Supplementary Material \cite{Suppl}.
In case of ppyH or 
bpy aromatic molecules, silver clusters are normally 
strongly bound to the N atoms of ppyH and bpy 
(\Fig{bpy-Agn-examples}a),  which is impossible
when interacting with 
IrPS where N atoms are oriented towards the central 
Iridium atom. Nevertheless, such structures have been also taken 
into account because we do not want to exclude the 
possibility of superposition of dispersion interactions. 
In structures where there is no
direct interaction between Ag and N atoms,
the silver cluster is bound to one of the rings 
(see \Fig{bpy-Agn-examples}b).
For the ppyH molecule, in most cases the cluster 
is attached to the heterocycle. Still another possibility is that 
one of the silver cluster's planes interacts 
with the $\pi$ aromatic system and
is approximately parallel to the plane of one or both aromatic rings
(see \Fig{bpy-Agn-examples}c, d). 

Upon complex formation the bond lengths
of the silver clusters, bpy, and ppyH
do not strongly change as compared to gas phase.
The most pronounced changes occur 
in ppyH and bpy torsional 
angles between the aromatic rings. The changes 
for bpy are in the range of 1-6\textdegree\,  
with several exceptions where 
these changes achieve values in the 
range 26-33\textdegree. For ppyH, there 
are even more cases where changes are in 
the range of 23-35\textdegree. 
This could be due to  the fact that  the phenyl-ring of 
ppyH is repelled once the silver cluster comes close 
to bind with the N atom. The distances 
between the closest atoms of the 
interacting subsystems  are 2.3-4.0 \AA.

\begin{figure}[t]
	\centering
\includegraphics[width=0.7\textwidth]{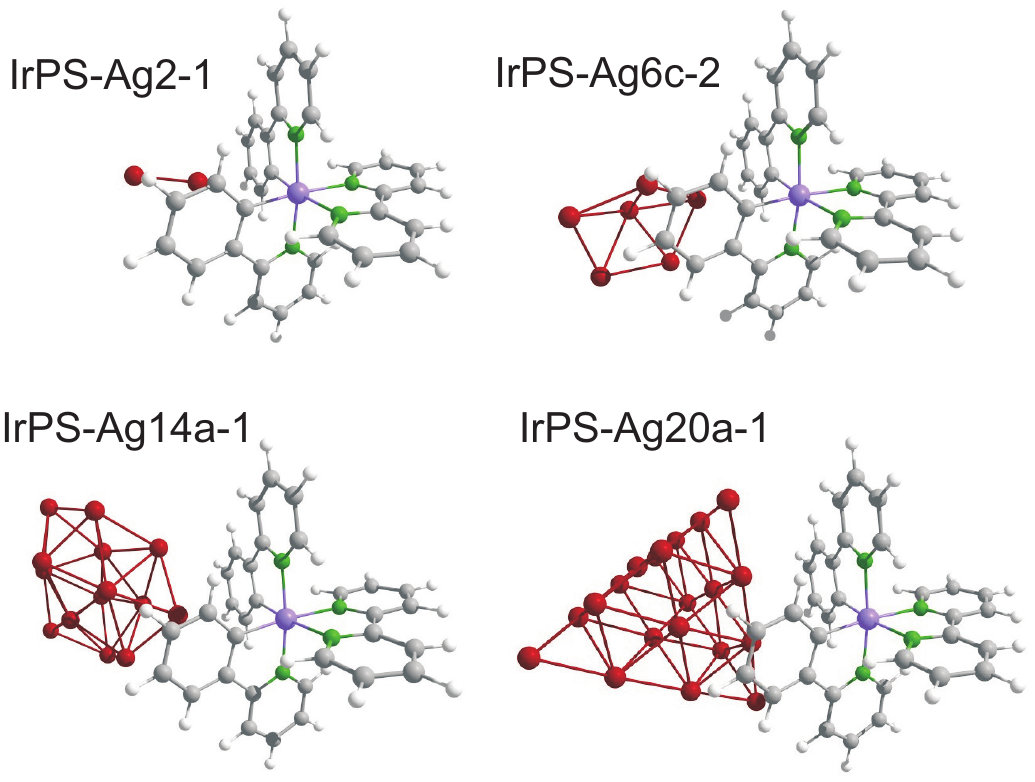}
\caption{\label{IrPS-Agn-examples}
Some examples of optimized \ce{IrPS-Ag_{$n$}} 
structures (for a full list  see Supplementary Material \cite{Suppl}).}
\end{figure}

In our previous study on the IrPS bound to small ($n\le 6$)
silver clusters \cite{Bokareva-pccp-2012}, we found
that configurations in which \ce{Ag_{$n$}} is 
situated in the cavities
between ligands are the lowest in energy. Note
that in all cases the interactions are "weak" and
no chemical bonds are formed. Here we  extend this study 
to \ce{IrPS-Ag_{$n$}} geometries up 
to 20 silver atoms, however, focussing on 
those structures where the cluster is located 
in the ''ppy-ppy'' cavity and $n$ is even. Similar to the 
small systems \cite{Bokareva-pccp-2012},
the geometry optimization was carried out without
symmetry constraints, except for some cases where 
the $\rm C_{2}$ point symmetry of the IrPS part
\cite{King-jacs-1987}
was retained. Some examples of  \ce{IrPS-Ag_{$n$}}
complexes can be found in \Fig{IrPS-Agn-examples}.

\subsection{Binding Energies}

Binding energies per silver atom of 
all structures (ppyH-Ag$_n$, 
\ce{bpy-Ag_{$n$}}, and \ce{IrPS-Ag_{$n$}}) 
selected for further
C$_{\rm 6}$ coefficient fitting 
are plotted in \Fig{Binding_all_eV} (details are given in the Supplementary Material \cite{Suppl}).
In general the range of binding energies 
(0.01-0.59 eV) is comparable
for all sets of model structures. 
This corresponds to physisorption
of the organic molecule on the silver 
clusters, with binding energies
being about 10 times smaller compared 
to binding energies of silver atoms 
within large nanoparticles \cite{Fernandez-prb-2004}.
From \Fig{Binding_all_eV} one can also notice
that $E_{\rm b}^{}$ decreases with increasing 
number of silver atoms in the system. 

Note that the RPA calculations were only
computationally affordable for structures 
containing up to 10 silver atoms
and bpy or ppyH but not IrPS as an organic counterpart.
Concerning the discrepancies between 
the PBE and MP2 calculations, it is observed that 
they are more pronounced for \ce{IrPS-Ag_{$n$}} 
systems, with MP2 results 
naturally being always larger
than those of PBE.
The RPA binding energies 
for \ce{bpy-Ag_{$n$}} and \ce{ppyH-Ag_{$n$}} 
lie in between PBE and MP2 ones, with RPA energies being in average
24\% lower that MP2. Next, these data on binding 
energies are used to optimize the dispersion coefficients.
\subsection{Problem-specific Dispersion Coefficients for Ir and Ag}
Using the QM:QM approach we obtained 
$C_{\rm 6}^{}$ coefficients for Ag
and Ir in two different ways as outlined 
in the Computational Details section.
The coefficients obtained either 
in two-steps or by simultaneous fitting 
as well as corresponding root mean squared deviations (RMSD)  
for binding energy fitting
are collected in Table~\ref{C6_coef}. 
For silver, the dispersion coefficient was
obtained using both RPA and MP2 as references.
The $C_{\rm 6}^{}$ coefficients used 
in the D2 model \cite{Grimme-jcc-2006}
are also given in Table~\ref{C6_coef}. 
It should be noted that
for Ag and Ir only estimates of coefficients
are available in D2. For example $C_{\rm 6}^{\rm Ag}$ 
was assumed to be an average of preceding group 
VIII and following group III element.
The corresponding $C_{\rm 6}^{}$ coefficients
of the D3 model are also included in the Table~\ref{C6_coef}
but two points should be highlighted before the 
comparison with the coefficients obtained in the current work.
First, they are coordination number dependent; in the table 
only values which were applied for the systems under investigation are collected.
Second, to fit all the dispersion forces in DFT-D3 
higher-order terms $C_{\rm 8}^{}$ and $C_{\rm10}^{}$
are also used.
We also included in Table~\ref{C6_coef} 
the values of $C_{\rm 6}^{}$ obtained within
the combination of dispersion-corrected density-functional
theory (the DFT+ van der Waals approach) 
\cite{Tkachenko-prl-2009}, with 
the Lifshitz-Zaremba-Kohn theory for 
the nonlocal Coulomb screening within the bulk \cite{Ruiz-prl-2012} .
The average value of $C_{6}^{}$ deduced from CCSD(T) interaction energies
between silver dimers at large distances 
is also provided in Table~\ref{C6_coef} \cite{Hatz-jpca-2012}.

Comparing new coefficients based on MP2 reference
and standard (D2) coefficients,
one notices that the present QM:QM $C_{\rm 6}^{}$ coefficients are 
substantially larger than those
of the standard D2 model, with the 
differences for Ag being more pronounced
than for Ir (increase
by 170-180\% vs. 25-70 \%). The two new 
$C_{\rm 6}^{\rm Ag}$ coefficients  are 
closer to each other than corresponding
$C_{\rm 6}^{\rm Ir}$ ones,  hinting at the minor
influence of Ir coefficient on the approximation
of $E_{\rm MP2}^{}-E_{\rm PBE}^{}$  differences
for the case of silver.

Here, one should take into account the possible deficiencies
of MP2 for predicting binding energies. For example, for benzene molecule
the  $C_{\rm 6}^{}$ coefficient evaluated by MP2 is overestimated by more than 40\% 
\cite{Tkachenko-jcp-2009}. These overestimations could be even
more pronounced for highly polarizable systems such as metal clusters and 
for periodic systems should lead to infinite 
$C_{\rm 6}^{}$ coefficient because of vanishing bandgap. 
Indeed, if we compare highly accurate $C_{\rm 6}^{}$ values from
Table~\ref{C6_coef} of Ref. \cite{Ruiz-prl-2012} for the bulk silver
within DFT-vdWsurf approach
which was obtained using the highly accurate experimental
dielectric function of the Ag bulk, our 
$C_{\rm 6}^{}$ is about one magnitude higher. 
\textcolor{black}{The $C_{6}^{}$ value for free silver atoms 
from DFT-vdW \cite{Tkachenko-prl-2009}
is about 3 times larger than for the bulk phase \cite{Ruiz-prl-2012}.} 
However, all our systems have
non-vanishing bandgap even up to \ce{Ag92}. For test cases 
(see the Supplementary Material), MP2
overestimates binding energies by only 10-40 \% if compared to CCSD.
The orbital shift in MP2 leads to almost linear scaling 
of energy \cite{Roos-cpl-1995} which
indirectly evidences the absence of problems with the bandgap. 
Nevertheless, if one compares the present results with
CCSD(T) for silver dimers, the
$C_{\rm 6}^{}$ coefficients obtain by MP2 are three
 times larger. Summarizing,
$C_{\rm 6}^{}$ coefficients obtained by MP2 are likely to
 be overestimated, but one can state that for small 
 systems they are larger than for the bulk metal.

Alternative to the MP2 reference we consider 
RI-RPA/def2-TZVP calculations. As noted before 
these have been performed for a smaller test set only, 
which did not include \ce{IrPS-Ag_{n}} structures and 
cases with more than 10 Ag atoms. Separate fitting leads to 
a substantially different $C_{6}^{\rm Ag}$ coefficient, which
is even smaller (by 28\%) than the corresponding 
standard D2 value, see Table~\ref{C6_coef}.
In order to investigate this point further we performed 
a test calculation (Test1 in Tab.~\ref{C6_coef}) where
the MP2 reference is taken for the same reduced set 
of structures. This yields a lowering of $C_{6}^{\rm Ag}$ 
with respect to the full MP2 set, but still a notably
 higher value (13\%) as compared with RPA. 

To scrutinize the large variability of 
$C_{\rm 6}^{\rm Ir}$  further two additional
fits have been performed. First, we 
fitted $C_{\rm 6}^{\rm Ag}$ using all sets of
test structures (MP2 reference) and assuming 
$C_{\mathrm 6}^{\mathrm Ir}=842.0$~eV $\AA ^{6}_{}$ (D2
model), and second,
we applied the standard D2 $C_{\mathrm 6}^{\rm Ag}$~=~255.69
and fitted $C_{\rm 6}^{\mathrm Ir}$ only for the \ce{IrPS-Ag_{$n$}} set
of structures.
The results are also shown in 
Table ~\ref{C6_coef} (entries ''Test2'', ''Test3'').
Taking $C_{\rm 6}^{\mathrm Ir}=842.0$~eV $\AA^{6}_{}$ 
leads to a slight increase of $C_{\rm 6}^{\mathrm Ag}$ 
if compared to the "simultaneous" value, with the
quality of fitting being practically the same. 
Setting $C_{\rm 6}^{\rm Ag}$
=255.69 led to an enormously increased $C_{\rm 6}^{\mathrm Ir}$, 
with the RMSD increasing dramatically as well. 
This finding can be rationalized as follows: 
The Ir atom is situated in the 
center of photosensitizer and hence 
screened by the ligands what hinders a direct
interaction with the silver clusters. To cover the same 
amount of dispersion with the standard fixed
value of $C_{\rm 6}^{\rm Ag}$ is only possible with an extremely large
coefficient for iridium. Hence, it can be argued that the actual coefficients
are extremely sensitive to the chemical environment, 
what in principle makes the simultaneous fit more reliable.
\textcolor{black}{The dependency of coefficients on environment hints that
in the current approach they are more akin to those of
DFT+vdW, XDM and more recent methods
like DFT+MBD, DFT+vdWsurf (for discussion see \cite{Bucko-prb-2013})
where screening effects are considered in a more complicated way.}

\textcolor{black}{Fitting the difference between PBE and a reference method (MP2 or RPA)
invokes naturally the question, which effects and contributions are actually included in
the resulting dispersion energy. DFT-D2 describes solely dipole-dipole interactions, 
whereas DFT-D3 additionally takes  triple-dipole interactions into account. MP2 
provides an approximate description of all vdW effects with moderate computational costs
but it is known to overestimate dispersion effects. In Ref. \cite{Tkachenko-jcp-2009}, a
dispersion-corrected MP2 version was suggested to
overcome this deficiency. Finally, as a consequence of its non-locality
the RPA approach  should describe the long-range vdW-interaction
in the most accurate way. Inclusion of dynamic electronic screening
extends the applicability of RPA to small-gap and metallic systems, which
cannot be treated generally  by MP2. }

\textcolor{black}{According to the original Grimme
DFT-D2 model, the dispersion contribution
is additionally multiplied by a damping function and a
global scaling factor $s_{6}^{}$ which is functional-dependent. 
In that way, the strength of dispersion interaction is adjusted for
different XC functionals. In case of PBE, $s_{6}=0.75$ is used. Such an ad hoc treatment  is 
 justified at short distances where it provides a correction to
the overestimated dispersion forces, but it has no 
direct meaning at long distances.
The Fermi-type damping function includes an 
additional empirical parameter $d$, the
steepness of the damping function which is universal 
for all functionals. As it was already
pointed out in Sec. \ref{sec:intro}, the damping function is 
needed to smoothly switch-off the dispersion
term at short distances. 
The choice of a particular type of damping function has no crucial
impact on the results (for a discussion see Ref. \cite{Grimme-jcp-2010}).
In more advanced DFT-D variants such as DFT-D3, TS, and XDM,
there is no direct scaling of $C_{6}^{}$ coefficients, 
but damping functions include more empirical parameters, like (order-dependent-)
coefficients which scale vdW (or cutoff) radii for each 
density functional applied. 
Furthermore, the cutoff radius itself is
also an arbitrary parameter that is not universal for different DFT-D approaches and 
should be chosen separately. 
In DFT-D3, the $s_{8}^{}$ parameter is used to scale the contribution of triple-dipole interactions and
$s_{6}^{}$-scaling is applied for  double-hybrid density functionals. 
Summarizing, all DFT-D schemes include empirical corrections justified by
physical meaning and by fitting to experimental or more precise
computational results. In the present treatment we do not aim at refitting the scaling factor and take the DFT-D2 model as developed previously. } 

In the following applications we continue to consider two sets of coefficients: First the MP2 result obtained from simultaneous fitting (PBE-D2*). This includes the complete set of test structures which is not available for the most likely more accurate RPA. Note, however, since no exact reference is available the judgement concerning accuracy of the two methods is solely based on results reported for other systems in literature. Therefore MP2 derived results will be contrasted to RPA ones which are supplemented by the $C_{6}^{\rm Ir}$ coefficient taken from the D2 model (PBE-D2**).

\begin{table}
\begin{center}
\begin{tabular}{c c c}
\hline
              \ \ & \ \  Ag   \ \ & \ \  Ir          \\
\hline
 D2 \cite{Grimme-jcc-2006} \ \ & \ \ 255.69 \ \ & \ \ 842.00 \\
 D3 \cite{Grimme-jcp-2010} \ \ & \ \ 160.53 (CN=1) \ \ & \ \ 182.60 (CN=3) \\
 Ruiz \cite{Ruiz-prl-2012} \ \ & \ \ 72.91 \textcolor{black}{(free 202.60)*} \ \ & \ \ -- \\
 Hatz \cite{Hatz-jpca-2012} \ \ & \ \ 216.96 \ \ & \ \ -- \\
 Separate (MP2) \ \ & \ \  690.41 (0.08)   \ \ & \ \  1420.60 (0.11)    \\                                                                                           
 Simultaneous (MP2) \ \ & \ \ 717.70 (0.09) \ \ & \ \ 1051.68 (0.09) \\
Separate  (RPA) \ \ & \ \ 183.18 (0.06) \ \ & \ \ --\\
 Test1 \ \ & \ \ 599.46 (0.05) \ \ & \ \ --\\
 Test2 \ \  & \ \ 727.92 (0.09) \ \ & \ \ 842.00 \\
 Test3 \ \ & \ \ 255.69 \ \ & \ \ 45667.49 (0.24)\\
  \hline
\end{tabular}	
\end{center}
\caption{\label{C6_coef}
Values of $ C_{\rm 6}^{\rm Ag}$ and $C_{\rm 6}^{\rm Ir}$
(in $\rm eV \AA^{6}_{}$) 
fitted separately and simultaneously with the QM:QM method
(RMSD of $E_{\mathrm ref}^{}-E_{\rm PBE-D*}^{}$
in eV$^{2}_{}$ are given in 
parenthesis) and from literature. For D3 value,
the coefficients are only given for coordination numbers (CN) 
in parenthesis. 
\textcolor{black}{*The unscreened (free) value from \cite{Tkachenko-prl-2009} is given for comparison.}
For the meaning of Tests see text.
}
\end{table}

\subsection{Applications}

First, the obtained coefficients for Ag and Ir were verified
for dissociation of two complexes: \ce{IrPS-Ag4} and
\ce{IrPS-Ag20}. The optimized (PBE/def2-SV(P)) geometries
were frozen and distances between Ir and closest
Ag atoms were changed. Single point calculations 
of obtained structures were done 
using PBE (pure and in different dispersion-included variants) and MP2 with the 
def2-SV(P) basis set.

\begin{figure}[t]
		\centering
\includegraphics[width=0.7\textwidth]{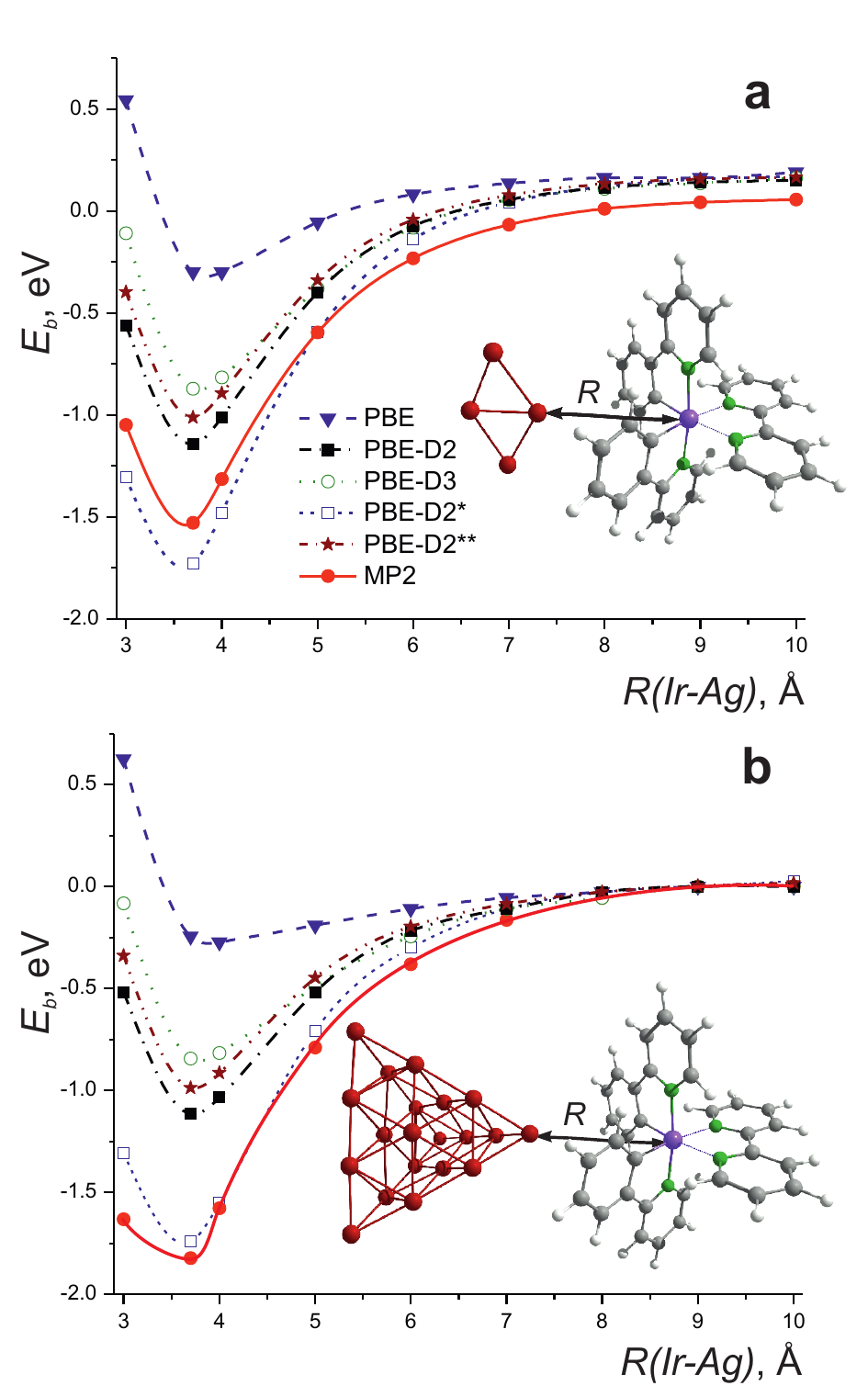}
\caption{\label{Diss_curves}
Dissociation curves of \ce {IrPS-Ag4} (a) and \ce {IrPS-Ag20} (b) complexes
calculated with MP2 and different PBE variants using the def2-SV(P) basis set. The legend for (b) part is
the same as for (a); connecting lines are just a guide to the eyes.
}
\end{figure}

The dependencies of calculated binding
energies on the Ir-Ag distance are plotted in \Fig{Diss_curves}.
Naturally, a very good agreement between MP2 and PBE-D2* curves
can be concluded for both complexes, with the D2, D3, and D2**
lying in the middle of MP2-PBE gap. 
Note that the minima of MP2 and 
DFT-D2* curves do not coincide since a 
larger basis set has been used for 
obtaining the C$_{\rm 6}$ coefficients. 
RPA based results as well as coefficients themselves
are lying between corresponding PBE-D2 and PBE-D3 curves,
demonstrating slightly slower binding energies
closer to dissociation limit. 

\begin{figure}[t]
		\centering
\includegraphics[width=0.7\textwidth]{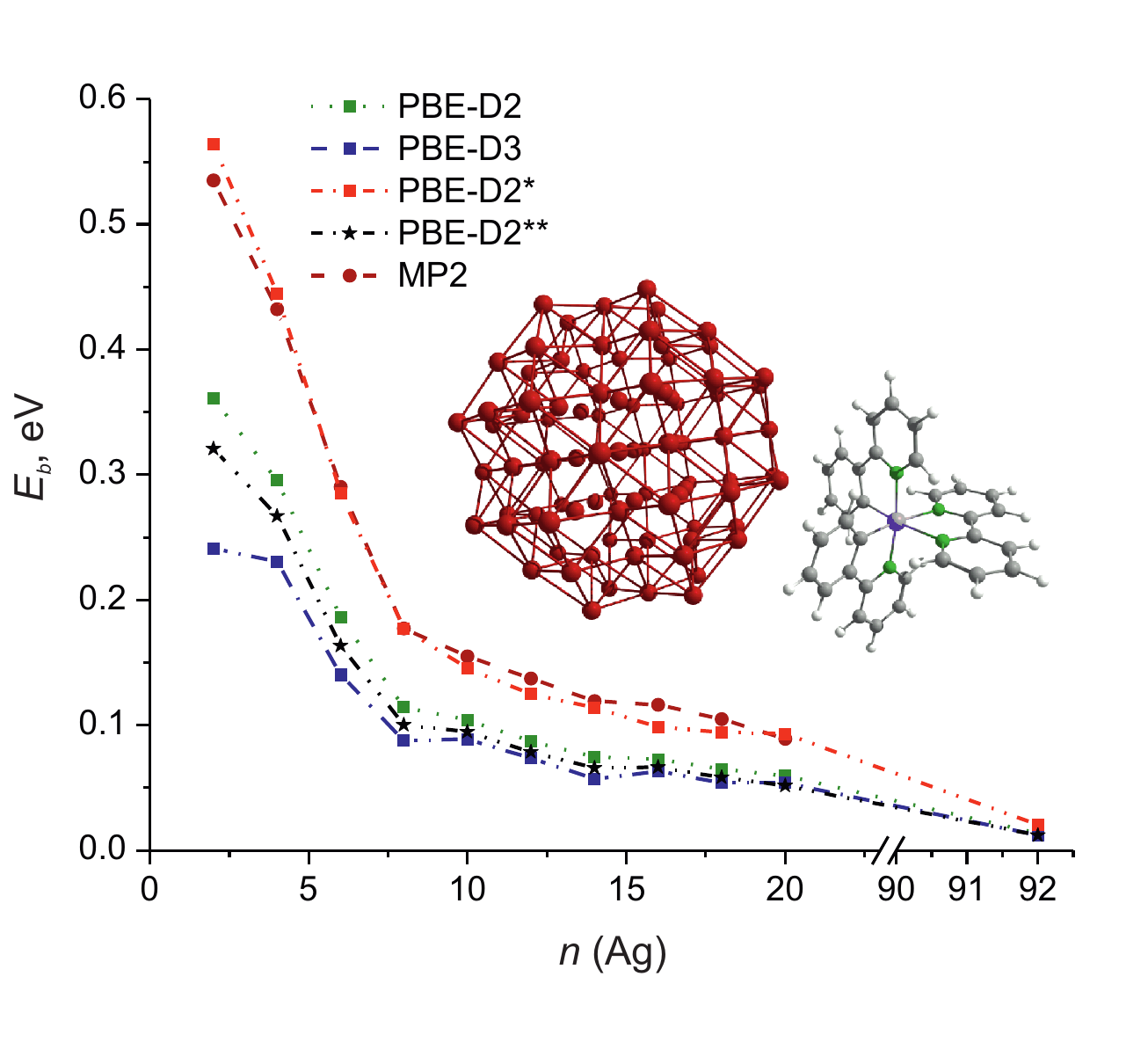}
\caption{\label{IrPS-Ag92}
Geometry of optimized complex \ce{IrPS-Ag92} as 
well as its binding energies evaluated with PBE-D
approaches compared to the MP2 binding energies (def2-TZVP basis set). For comparison binding energies for smaller systems are given as well.
}
\end{figure}

As a second application, we evaluated the binding 
energy of the large nearly
spherical silver cluster \ce{Ag92}.
The initial geometry of this cluster was cut 
from the fcc bulk silver and optimized 
with PBE/def2-SV(P) separately. Then the silver 
cluster was placed
in the ppy-ppy cavity of the photosensitizer 
analogous to the smaller
systems. The constructed
complex was partially optimized using PBE-D2*, 
i.e. only those silver atoms nearest
to the IrPS and all atoms
of IrPS were allowed to relax upon optimization. 
(Note for the purposes of comparison,
we also performed an optimization with PBE-D2 and PBE-D3;
the changes in geometries are only minor (up
to 0.06 \AA{} and 0.2\textdegree)
and can be neglected.)
The resulting geometry is shown in \Fig{IrPS-Ag92} together
with the corresponding binding energies
per one silver atom
calculated within different PBE-D models. The \ce{IrPS-Ag92}
complex has no symmetry
and the distance between central Ir atom and
nearest silver atom is 5.2 \r{A} which 
is slightly higher than in the case of small silver 
clusters. Due to
steric reasons the large silver cluster can 
not come closer without significant distortions
in its shape. The binding energy of the 
largest investigated system per
one silver atom is very small and  in 
the region of 0.004--0.021 eV for all PBE-D
variants. Similar to the small systems, the account 
for dispersion forces
increases the binding energy $E_{\rm b}^{}$: from  0.004 (PBE) 
to 0.012 (both PBE-D2 and PBE-D3, PBE-D2**) and
to 0.021 eV (for PBE-D2*). The obtained value 
also confirms the decrease of $E_{\rm b}^{}$ per
1 silver atom with increasing system size. 
The trend can be clearly seen in \Fig{IrPS-Ag92}
where binding energies for selected examples of \ce{IrPS-Ag_{n}^{}}
calculated with MP2 and all the PBE-D variants (within def2-TZVP basis set)
are plotted. For clarity, only cases with 
the largest binding energies within MP2 are selected.

\begin{figure}[t]
		\centering
\includegraphics[width=0.7\textwidth]{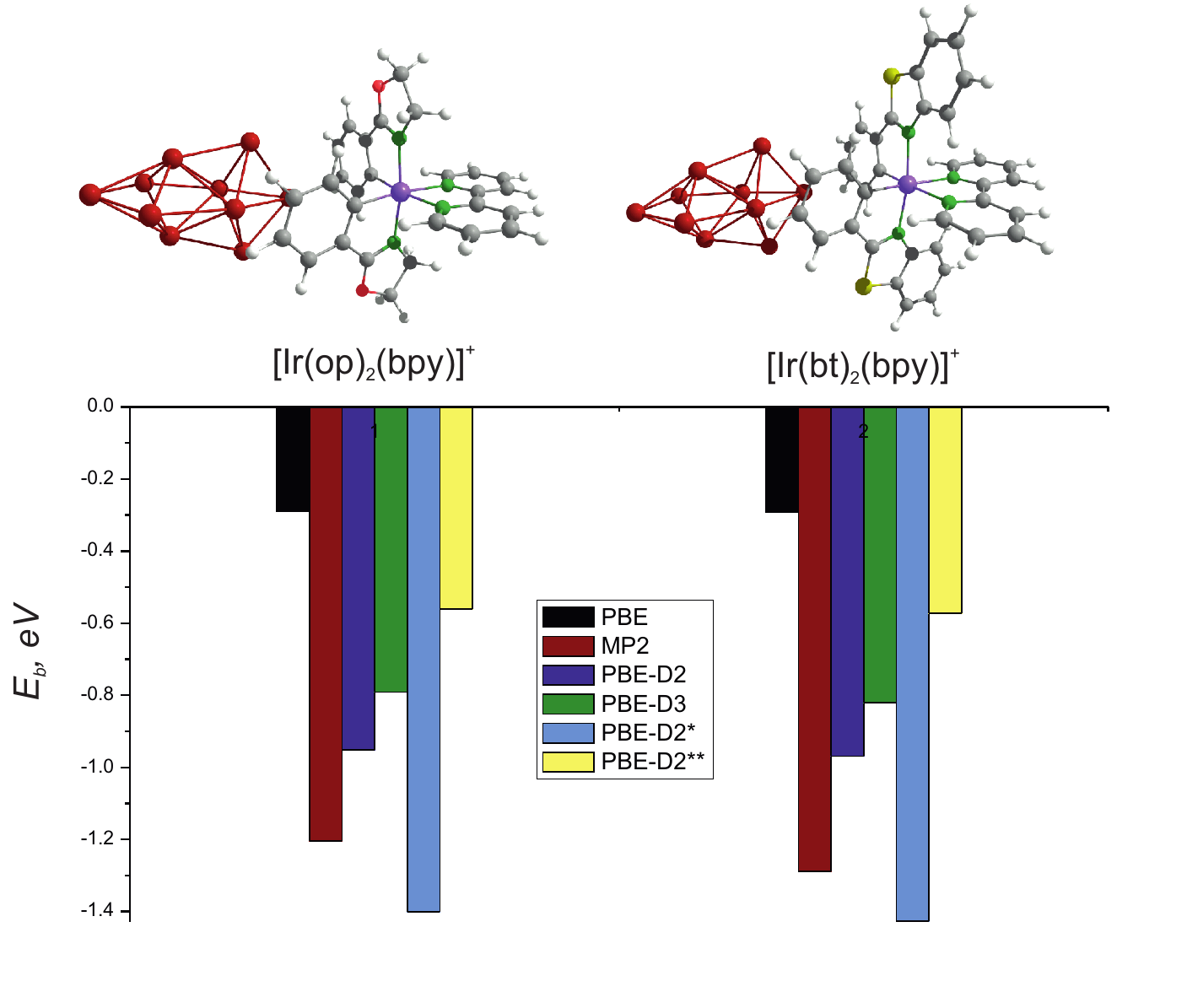}
\caption{\label{Derivatives}
Geometries and comparison of binding energies 
of two further photosensitizers containing O or S atoms using the def2-SV(P) basis set.
}
\end{figure}

Finally, we apply the new coefficients to derivatives 
of IrPS. Although our coefficients are by construction
suited for the particular photosensitizer with ppy and bpy
ligands (isolated or located around Ir central atom)
we wanted to scrutinise the sensitivity with respect 
to the type of heteroatom. For this reason we investigated 
the binding between and
\ce{Ag10} cluster an a photosensitizers
containing oxygen (\ce{[Ir(op)2(bpy)]+}) 
and sulphur (\ce{[Ir(bt)2(bpy)]+}) in the 
ligands. The structures of optimized complexes
are depicted in \Fig{Derivatives}
and they are overall similar to those of IrPS.
Again the silver cluster is located in the cavities
between the two ligands, with the distance
between the central Ir atom and the
nearest silver atom being about 4.8--4.9 \r{A}. The 
binding energies are about 1.4 eV  (or 0.14 eV
per one silver atom) for both cases,
which is only slightly lower than those of \ce{IrPS-Ag10}
and corresponds again to a weak interaction 
(physisorption). 

%
\section{Conclusions}
\label{sec:Conc}
%
The interaction of Ir(III) photosensitizers containing
ppy and bpy ligands with small silver clusters
Ag$_n$ ($n$=2--20) is studied using dispersion-corrected
density functional theory together with the RPA and  MP2 methods.
The goal has been to develop a system-specific 
set of $C_{\rm 6}^{}$ parameters for Ir
and Ag atoms in the spirit of the D2 correction. 
To this end the QM:QM scheme was employed 
for a set of model structures. 
An important aspect for this particular type of system
is that the Ir atom is shielded by the ligands. 
As a consequence the Ir parameters turn out to 
be rather sensitive to the actual method of fitting, 
whereas the Ag coefficients are more robust. 

In general binding of silver clusters to IrPS is weak 
and in the physisorption range. The binding energy 
per silver atom decreases with the 
size of the cluster to become as small as 0.01--0.02 eV
for the largest cluster studied ($n=92$).

Although our new coefficients are by construction
suited for the particular example of interaction 
of silver clusters with organic or metalorganic 
molecules containing ppy and bpy ligands, 
the transferability of new $C_{\rm 6}^{}$ 
coefficients for similar systems
was shown using the exemplary cases 
of IrPS containing O and S heteroatoms.
 
In order to derive $C_{\rm 6}^{}$ two references have been considered, i.e. MP2 and RPA. It was found that MP2 did not give convergence  and bandgap-related problems, most likely since the considered metal clusters still have molecular character and can be described with respective orbitals. However, based on reports in literature one would expect that RPA is more reliable for the description of binding in these weakly bound organic/inorganic hybrid systems. Therefore, the $C_{\rm 6}^{\rm Ag}$ value obtained on the RPA level should be considered as being more accurate compared to MP2. This value turns out to be close to the ones of the Grimme D2/D3 sets. However, this similarity should be taken with care. In view of the fact that the polarizability is determined  by a non-local response kernel, the reduction to a single coefficient for an arbitrary arrangement of atoms is quite an approximation. As an additional caveat we should mention that, of course, extraction of dispersion coefficients in the present QM:QM scheme assumes that other contributions to the interaction do not differ between the QM methods. This is not necessarily the case and differences might result from electrostatic or induction contributions. Therefore, in view of the size of the systems considered here, crucial tests against experimental data will be required for further validation.

\section*{Acknowledgements}
This work has been financially suppor\-ted by 
the ESF project ''Nanostructured Materials 
for Hydrogen Production (Nano4-Hydrogen)'' and the BMBF project ''Light2Hydrogen'' (''Spitzenforschung und 
Innovation in den Neuen L\"andern'').


\clearpage\newpage

\clearpage\newpage

\clearpage\newpage

\clearpage\newpage

\begin{figure}[t]
	\centering
\includegraphics[width=0.5\textwidth]{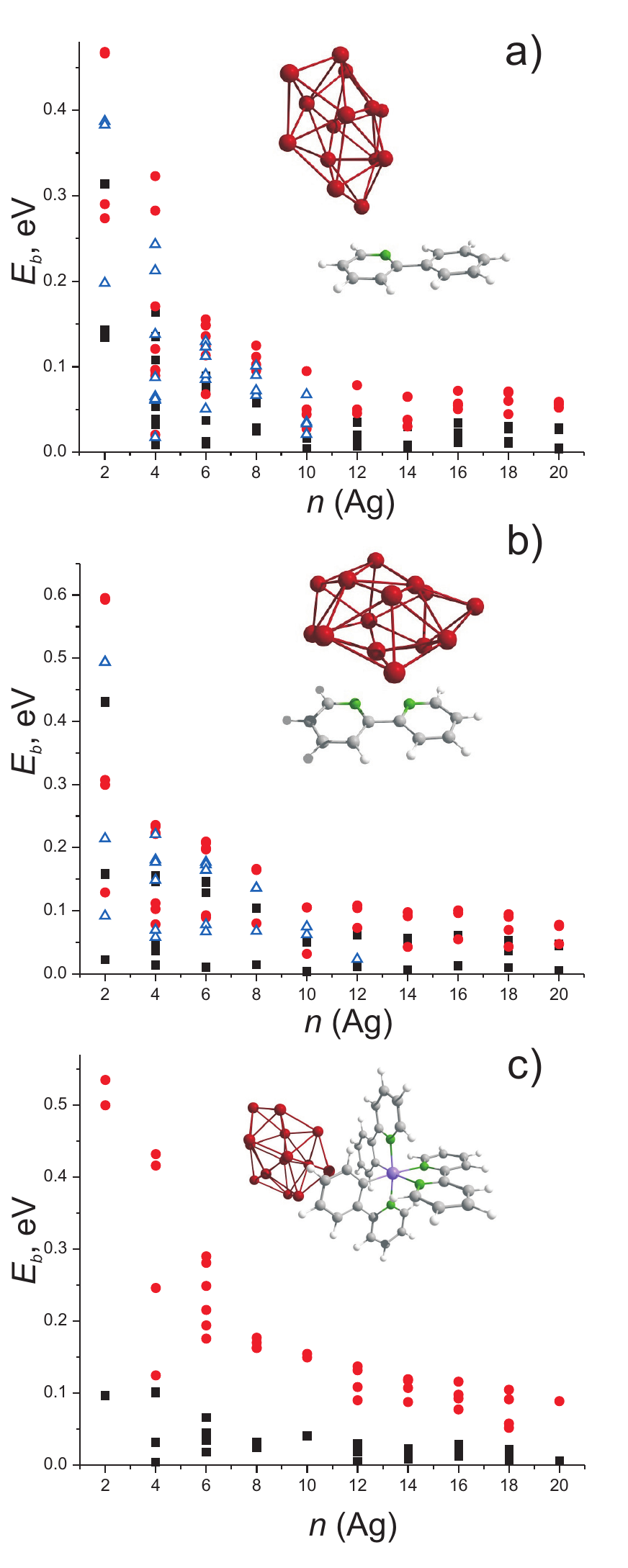}
\caption{\label{Binding_all_eV}
Binding energies per silver atom of all model structures under
study: a) \ce{ppyH-Ag_{$n$}}, b) \ce{bpy-Ag_{$n$}}, c)
\ce{IrPS-Ag_{$n$}}. Black squares: PBE, blue hollow triangles: RPA,
red circles: MP2. 
For illustration some examples for \ce{Ag14} are shown. Notice that for simplicity, we do not mark the corresponding points (PBE, RPA, MP2) for each structure.}
\end{figure}

\clearpage\newpage

\clearpage\newpage

\clearpage\newpage

\end{document}